\documentclass[11pt]{article}
\usepackage{amssymb}
\usepackage{amsmath}
\usepackage{graphicx}
\newcommand{\newtext}[1]{\text{#1}}

\def\beq{\begin{eqnarray}}    
\def\eeq{\end{eqnarray}}      
\newcommand{\be}{\begin{equation}}
\newcommand{\ee}{\end{equation}}
\newcommand{\bea}{\begin{eqnarray}}
\newcommand{\eea}{\end{eqnarray}}
\newcommand{\lb}{\label}

\newcommand{\OMo}{\Omega_m^0}

\newcommand{\ORo}{\Omega_{r}^0}

\newcommand{\OLo}{\Omega_{\Lambda}^0}

\newcommand{\rc}{\rho_c}
\newcommand{\rco}{\rho_{c}^0}
\newcommand{\rmo}{\rho_{m}^0}

\newcommand{\rM}{\rho_m}
\newcommand{\rmr}{\rho_m}
\newcommand{\pmr}{p_m}
\newcommand{\rMo}{\rho_{m}^0}

\newcommand{\rR}{\rho_r}

\newcommand{\wm}{\omega_m}

\newcommand{\rL}{\rho_{\CC}}

\newcommand{\rLo}{\rho_{\CC}^0}

\newcommand{\CC}{\Lambda}

\newcommand{\CH}{C_H}
\newcommand{\CHd}{C_{\dot{H}}}
\newcommand{\Hd}{\dot{H}}
\newcommand{\bnu}{\bar{\nu}}
\newcommand{\rDE}{\rho_{\rm DE}}

\newcommand{\nueff}{\nu_{\rm eff}}
\newcommand{\xiM}{\xi_m}
\newcommand{\xiR}{\xi_r}

\newcommand{\rRo}{\rho_r^0}
\newcommand{\rDent}{\rho_{\rm DE}^{\rm entr}}

\newcommand{\mysection}[1]{\section{#1}
\renewcommand{\theequation}{\thesection.\arabic{equation}}
\setcounter{equation}{0}}

\newcommand{\mysubsection}[1]{\subsection{#1}
\renewcommand{\theequation}{\thesubsection.\arabic{equation}}
\setcounter{equation}{0}}

\begin{document}



 \hyphenation{cos-mo-lo-gi-cal
sig-ni-fi-cant par-ti-cu-lar}




\begin{center}
{\large \textsc{Generalizing the running vacuum energy model and comparing with\\
the entropic-force models}} \vskip 2mm
%
%
 \vskip 8mm

\textbf{Spyros Basilakos$^1$, David Polarski$^{2,3}$, Joan
Sol\`{a}$^{4,5}$}
%
%

\vskip0.5cm $^1$~Academy of Athens, Center for Astronomy and Applied
Mathematics, Athens, Greece

\vskip0.5cm

$^2$~Universit\'e Montpellier 2, Lab. Charles Coulomb UMR 5221, F-34095 Montpellier, France \\
$^3$~CNRS, Laboratoire Charles Coulomb UMR 5221, F-34095 Montpellier,
France

\vskip0.5cm

$^4$~High Energy Physics Group, Dept. ECM, Univ. de Barcelona, Av.
Diagonal 647, E-08028 Barcelona, Catalonia, Spain

$^5$~Institut de Ci{\`e}ncies del Cosmos (ICC),  Univ. de Barcelona, \\
Av. Diagonal 647, E-08028 Barcelona, Catalonia, Spain


\vskip0.5cm

E-mails: svasil@academyofathens.gr, david.polarski@univ-montp2.fr,
sola@ecm.ub.es \vskip2mm

\end{center}
\vskip 15mm

\begin{quotation}
\noindent {\large\it \underline{Abstract}}.\ \
We generalize the previously proposed running vacuum energy model by
including a term proportional to ${\dot H}$, in addition to the existing
$H^2$ term. We show that the added degree of freedom is very constrained
if both low redshift and high redshift data are taken into account.
Best-fit models are undistinguishable from $\Lambda$CDM at the present
time, but could be distinguished in the future with very accurate data at
both low and high redshifts.
We stress the formal analogy at the phenomenological level of the running
vacuum models with recently proposed dark energy models based on the
holographic or entropic point of view, where a combination of ${\dot H}$
and $H^2$ term is also present.
However those particular entropic formulations which do not have a constant
term in the Friedmann equations are not viable.
The presence of this term is necessary in order to allow for
a transition from a decelerated to an accelerated expansion.
In contrast, the running vacuum models, \newtext{both the original} and
the generalized one introduced here contain this constant term in a more
natural way. Finally, important conceptual issues common to all these
models are emphasized.
\end{quotation}
\vskip 8mm


\newpage

\vskip 6mm

 \noindent \mysection{Introduction}
 \label{Introduction}



The longstanding dark energy (DE) problem was originally presented in the
form of the cosmological constant (CC) problem\,\cite{CCproblem}.
Whichever way it is formulated, the CC problem appears as a tough
polyhedric conundrum which involves many faces: not only the problem of
understanding the tiny current value of the DE density $\rDE$ in the
context of quantum field theory (QFT) or string theory, but also the
cosmic coincidence problem, i.e. why the density of matter $\rM$ is now so
close to $\rDE$. Dynamical DE models are helpful in order to improve the
situation. They can appear in different formulations of fundamental
physics. Popular possibilities are, among others, quintessence and phantom
energy in its various forms\,\cite{quintessence}, and scalar-tensor
models\,\cite{ScalarTensor}. Furthermore, modified gravity is another very
interesting option, which has been intensively explored in the recent
literature, see e.g.\,\cite{ModGrav1,ModGrav2,Relax}.

But a class of cosmic accelerating models which we wish to explore in this
paper is that of dynamical vacuum energy models. They have been proposed
since long ago -- see e.g.\, \cite{MiniReview11,Fossil07,SS09} and
references therein. Some of these ``running'' vacuum models are a possible
clue for tackling one or more aspects of the CC problem. Despite the
various phenomenological existing studies of time evolving vacuum
models\,\cite{OldLambdaTime}, some of them are expected on more
fundamental grounds, e.g. within the context of QFT in curved
space-time\,\cite{Fossil07,SS09}. In fact, it is difficult to conceive an
expanding universe with a strictly constant value of the vacuum energy
density $\rL=\CC/(8\pi\,G)$, namely one that has remained unchanged since
the origin of time. It is much more natural to expect that the vacuum
energy is a dynamical quantity as the universe itself, and thereby
sensitive to time evolving functions such as the the Hubble rate $H=H(t)$
or the scale factor $a=a(t)=(1+z)^{-1}$ ($a_0=1$). In these models, the
need for scalars is obviated and nevertheless a phenomenologically viable
description for the dynamical nature of the vacuum energy is achieved. Not
only so, some of these models have been successfully tested against the
latest cosmological data, see e.g. the recent
studies\,\cite{BPS09,GSBP11}. Remarkably, some particular formulations of
them have been used to improve both the cosmic coincidence
problem\,\cite{LXCDM12} and the tough ``old CC problem'', i.e. the fine
tuning problem -- see e.g. the recent attempts within the context of
modified gravity\,\cite{Relax}.

More recently, Verlinde\,\cite{Verlinde10} proposed that the gravitational
field equations can be derived from the second law of thermodynamics in a
way that would render the gravity force quite literally as a kind of
``entropic force'' (which is certainly not the case in e.g.
Jacobson's\,\cite{Jacobson95} and Padmanabhan's
approaches\,\cite{Padmanabhan10}, in which the entropic formulation is much
more general). When Verlinde's entropic version is applied to cosmology,
the DE does not exist anymore as such, but is mimicked in an effective way
by the acceleration associated to the entropic force acting outwards the
cosmic horizon. It is this particular formulation that can be called
``entropic-force cosmology'' which was first explored in
Ref.\,\cite{Easson10,Easson10b}\, and later on by various authors -- see
e.g. \,\cite{Casadio10,Koivisto10,Entropic-others}.
%
%
We emphasize that these models seem to lead to the same effective
Friedmann equations as the aforementioned running vacuum
models\,\cite{MiniReview11,Fossil07,SS09} with the notable difference that
some of these entropic-force models, but not necessarily all of its
versions, do not yield a constant term in their Friedmann equations.

The plan of the paper is as follows. In Section 2 we review the running
vacuum model followed by a comparative discussion with the entropic-force
models, and we emphasize the analogy at the level of the equations of
motion. In Section 3 we present the background cosmology for these models.
We show that the entropic-force cosmology appears as a particular case of
the generalized running vacuum model. After comparing and fitting them to
the data in Section \ref{sect:fitting}, we provide our discussion and
final conclusions in Section \ref{sect:conclusions}.

\mysection{Running vacuum energy and entropic-force models}
\label{sect:runninentropic}

As mentioned in the introduction, dynamical dark energy is an attractive
possibility in order to explain certain aspects of the cosmological
constant problem. In this section we review the idea of running vacuum
energy, which was suggested in the literature long ago\,\cite{SS-old}, and
we take opportunity to compare it with the more recent notion of entropic
dark energy, specially some recent formulations of it\,\cite{Easson10,
Easson10b}. They are formally similar but present also important
differences which lead to significant phenomenological implications.  The
latter will be analyzed in subsequent sections.

\mysubsection{Running vacuum energy as dynamical dark energy}
\label{sect:runningv}

The running vacuum energy in QFT in curved space-time derives from the
renormalization group (RG) equation suggested in the literature for $\rL$
-- see \cite{MiniReview11} and references therein:
\begin{equation}\label{seriesLambda}
\frac{d\rL(\mu)}{d\ln\mu^2}=\frac{1}{(4\pi)^2}\left[\sum_{i}\,B_{i}M_{i}^{2}\,\mu^{2}
+\sum_{i}
\,C_{i}\,\mu^{4}+\sum_{i}\frac{\,D_{i}}{M_{i}^{2}}\,\mu^{6}\,\,+...\right]
\equiv \,n_2\,\mu^2+{\cal O}(\mu^4)\,,
\end{equation}
where $M_{i}$ are the masses of the particles contributing in the loops,
and $B_{i},C_i,..$ are dimensionless parameters. The equation
(\ref{seriesLambda}) gives the rate of change of the quantum effects on
the CC as a function of the scale $\mu$. Only the ``soft-decoupling''
terms of the form $\sim M_i^2\,\mu^2$ remain in practice, as the $M_i^4$
ones would trigger a too fast running of the cosmological
term\,\footnote{The main contribution to the running of $\rL$ clearly
comes from the heaviest fields in a typical GUT near the Planck scale,
i.e. those with masses $M_i\sim M_X\lesssim M_P$. See e.g. \cite{Fossil07}
for a specific scenario within this class of models, where the one-loop
contribution for the $B_i$ coefficients  is explicitly given.}. The
approximate integrated form of (\ref{seriesLambda}) is very simple:
\begin{equation}\label{GeneralPS}
\rL(H)=n_{0}+n_{2}H^{2}+{\cal O}(H^4)\,,
\end{equation}
where, following the aforesaid works, we have set $\mu=H$ as the
characteristic mass scale for FLRW-like universes, and we will neglect the
(much) smaller higher order powers of $H$. Indeed, notice that only even
powers are allowed by the general covariance, and hence no other
$H^{2n}$-terms beyond $H^2$ (not even $H^4$) can contribute significantly
on the \textit{r.h.s.} of equation (\ref{GeneralPS}) at any stage of the
cosmological history below the GUT scale $M_X\lesssim M_P$, so that we
omit them.  The additive constant term $n_0$ in (\ref{GeneralPS}) appears
in a natural way in this framework upon integrating the RG equation. It
will play a fundamental role in our discussion. Both $n_0$ and $n_1$
become related by the boundary condition $\rL(H_0)=\rLo$, which is to be
satisfied by (\ref{GeneralPS}) at present, $H_0=H(t_0)$. As a result these
coefficients can be conveniently rewritten as follows:
\begin{equation}\label{n0n2}
n_0=\rLo-\frac{3\nu}{8\pi}\,M_P^2\,H_0^2\,,\ \ \ \ \
n_2=\frac{3\nu}{8\pi}\,M_P^2\,,
\end{equation}
where from (\ref{seriesLambda}) we have defined the important
dimensionless parameter
\begin{equation}\label{nu1}
\nu=\frac{1}{6\pi}\, \sum_i B_i\frac{M_i^2}{M_P^2}\,.
\end{equation}
This parameter provides the main coefficient of the $\beta$-function for
the running of the vacuum energy. The coefficients $B_i$ in (\ref{nu1})
can be computed from the quantum loop contributions of fields with masses
$M_i$, and hence $\nu$ is naturally expected to be non-vanishing and small
($|\nu|\ll 1$). For instance, for GUT fields with masses $M_i$ near
$M_X\sim 10^{16}$ GeV, a natural estimate lies in the approximate range
$\nu=10^{-5}-10^{-3}$ \,\cite{Fossil07}. As a result we also expect a mild
running of $\rL$, hence a dynamical DE framework which is healthfully
close to the well tested concordance $\CC$CDM model in which $\rL$ is
strictly constant. This particular situation is retrieved only for
$\nu=0$, for which $\rL=\rLo$ at all times. However, there is no obvious
reason for $\nu$ to be strictly vanishing in QFT in curved space-time.
Therefore in the general case we should have a time evolution law for the
vacuum energy (\ref{GeneralPS}), whose leading contribution can be
presented as follows:

\begin{equation}\label{RGlaw2}
 \rL(H)=\rLo+ \frac{3\nu}{8\pi}\,M_P^2\,(H^{2}-H_0^2)\,.
\end{equation}
Substituting (\ref{RGlaw2}) in the general acceleration law for a
FLRW-like universe in the presence of a vacuum energy density $\rL$, we
find
\begin{equation}\label{vacuuma}
\frac{\ddot{a}}{a}=-\frac{4\pi\,G}{3}\,(\rmr+3\pmr-2\rL)=
-\frac{4\pi\,G}{3}\,(1+3\wm)\,\rmr+C_0+\nu\,H^2\,,
\end{equation}
with
\begin{equation}\label{C0}
C_0=\frac{8\pi G}{3}\,\rLo-\nu\,H_0^2=\frac{\CC}{3}-\nu\,H_0^2=H_0^2\,\left(\OLo-\nu\right)\,.
\end{equation}
Here $\wm=\pmr/\rmr$ is the equation of state (EoS) for a generic
component of matter ($\wm=0$ and $1/3$ for non-relativistic and
relativistic matter, respectively), and $\OLo=\CC/(3H_0^2)$ is the
cosmological CC parameter whose observational value is $\OLo\simeq 0.73$.
We note the presence of the constant term $C_0\propto n_0$. As warned
before, this term will play an important role in our study.

\mysubsection{Entropic-force models and effective dark energy}
\label{sect:entropicforce}

It is interesting that an effective dynamical dark energy component
similar to the one derived in the previous section can also be motivated
within the context of the entropic models. In a particular version of this
framework, called the entropic-force models\,\cite{Verlinde10}, the
holographic screen is thought to induce a force $\textbf{F}=T\,\nabla S$
on a test particle near the screen, where $T$ is the temperature of the
screen and $\nabla S$ is the change of entropy associated with the
information contained in it (which involves a large number of d.o.f.). The
screen is supposed to increase its entropy when the test particle
approaches it. Therefore, $\nabla S$ and the normal ${\bf n}$ on the
screen (pointing towards the particle, located in the inner volume bounded
by the screen) have opposite signs. Since the force is directed towards
the screen we have $F=-T dS/dr$, with $dr$ the distance of the nearby
particle to the screen. When applied to cosmology\,\cite{Easson10}, the
entropy of the Hubble horizon $R_H=c/H$ is obtained from Bekenstein's
formula $S_H=A_H\,k_B/4l_P^2$, where $A_H=4\pi R_H^2$ is the area of the
horizon and $l^2_P=G\hbar/c^3$ is the Planck's length squared\,\footnote{
Note that for the sake of better clarity, we keep $\hbar$ and $c$ in this
section, but natural units $\hbar=1=c$ for the rest.}. The change of
entropy when the radius of the horizon increases by $dr$ is simply
$dS_H=2\pi\,(R_H\,k_B/l_P^2)\,dr$. Inserting it in the formula for the
pressure exerted by the entropic force on the cosmological expansion,
$P=F/A=-(T/A)\,dS_H/dr$, and estimating that the horizon temperature is
$T=(\hbar/k_B)(H/2\pi)$ (proportional to the de Sitter
temperature)\,\cite{Easson10}, one finally obtains $P=-(2/3)\,\rc \,c^2$,
where $\rc=3H^2/(8\pi G)$ is the critical density. The minus sign in the
pressure is of course the characteristic feature of the accelerated
expansion in this entropic version. Apart from some coefficients that
depend on the estimations made, and which are not essential for the
argument, the basic result is that $P\propto -\rDent$, where $\rDent\sim
H^2\,M_P^2$ (with $M_P=G^{-1/2}$) is the quantity that plays the role of
effective DE in this entropic model. This framework suggests that the
entropic force leads to an effective DE density which is dynamical: it
specifically evolves as the square of the Hubble rate\,\footnote{Apart
from the running vacuum model of sect. \ref{sect:runningv}, other
frameworks involving the dynamical term $\sim H^2$ (treated as the full DE
density or as a component of it) were suggested
in\,\cite{Padmanabhan:2004qc} and more recently in\,\cite{Maggiore-Bilic},
all of them involving the idea of vacuum fluctuations.}.  By Friedmann's
equation, it immediately follows that at the present time the value of
$\rDent$ would be predicted in the ballpark of the measured vacuum energy
density: $\rDent(t_0)\sim H_0^2 M_P^2\sim\rLo\sim 10^{-47}$ GeV$^4$, where
$\rLo=\Lambda/(8\pi G)$.

Since the previous (entropic inspired) result is essentially a surface
effect from the horizon, one may think of fully generalizing it by
considering the gravitational action for space-times with
boundaries\,\cite{Hawking96}. This is achieved by adding the boundary
action term $I_B$ to the standard Einstein-Hilbert action, $I_{EH}$,
namely:
\begin{equation}\label{IB}
I_{EH}+I_B=\frac{1}{16\pi G}\int_{{\cal M}}\,d^4
x\sqrt{|g|}\,R\,+\,\frac{1}{8\pi G}\int_{\partial{\cal M}}\,d^3
y\sqrt{|h|}\,K\,.
\end{equation}
Here $h$ is the determinant of the metric $h_{ab}$ on the boundary
$\partial{\cal M}$, induced by the bulk metric $g_{\mu\nu}$ of ${\cal M}$,
and $y^a$ are the coordinates on $\partial{\cal M}$. Furthermore, $K$ is
the trace of the second fundamental form (or extrinsic curvature); if
$n^{\mu}$ is the normal on the boundary, it can be written as
$K=\nabla_{\mu}n^{\mu}$. The complete action is $I=I_{EH}+I_{B}+I_m$,
where $I_m$ represents the ordinary matter contribution. As a mere
technicality, let us point out that the precise definition of the boundary
term $I_B$ should actually include an overall sign, which is plus or minus
depending on whether the hypersurface $\partial{\cal M}$ is space-like
($n^{\mu}n_{\mu}=+1$) or time-like ($n^{\mu}n_{\mu}=-1$), respectively. We
exclude null surfaces for this consideration. Notice that the precise
coefficient in front of the boundary integral $I_{B}$ is chosen in such a
way that the surface terms generated from the metric variation of $I_{EH}$
are exactly canceled by the metric variation of $I_{B}$, provided the
variation $\delta g^{\mu\nu}$ is performed in such a way that it vanishes
on $\partial{\cal M}$, i.e. provided the induced metric $h_{ab}$ on the
boundary is held fixed. It follows that, in the presence of $I_B$, the
standard form of Einstein's equations is preserved even if the space-time
has boundaries.

The authors of Ref.\,\cite{Easson10} presumably used the above
interpretation of $I_B$ as a way to generalize the entropic force argument
given before, in the following way. As the surface terms emerging from the
variation of $I_{EH}$ are canceled by $\delta I_B$, they assumed that if
the total action would not contain $I_B$ the contribution of the
aforementioned surface terms to the field equations would be of the order
of the effect induced on them by $I_B$, estimated as $R$ times the
prefactor $1/(8\pi G)$ in $I_B$, i.e. $(12 H^2+6\dot{H})/(8\pi G)$ --
evaluated in the FLRW metric, in which $H=\dot{a}/a$ and $\dot{H}=dH/dt$.
However, since this is probably just a rough estimate of the effect, they
finally proposed to generalize the corresponding acceleration equation for
the scale factor in the form:
\begin{equation}\label{entropica}
\frac{\ddot{a}}{a}=-\frac{4\pi\,G}{3}\,(1+3\wm)\,\rmr+\CH\,H^2+\CHd\,\Hd\,.
\end{equation}
However not all of the models considered in \cite{Easson10} are of this
type \footnote{G. Smoot, private communication}.
The new ingredients are $\CH$ and $\CHd$, which are certain (presumably
small) dimensionless coefficients to be fitted to the observational data.
Let us also mention that there can be higher order quantum corrections on
the \textit{r.h.s.} of Eq.\,(\ref{entropica}) -- cf. \,\cite{Easson10b}.
We have neglected these effects for the present discussion because they
have no impact for virtually any time in the history of the universe after
inflation. This is in line with our approximation of ignoring the ${\cal
O}(H^4)$ quantum corrections also in the running vacuum model discussed
before.

Let us point out that the field equations (\ref{entropica}) are not
necessary derived from a fundamental action. Let us recall that in the
most general entropic-holographic formulations, gravity is conceived as an
emergent phenomenon\,\cite{Padmanabhan10}, and in this sense the
gravitational field equations need not necessarily be deducible from a
fundamental action at the present macroscopic level of description, even
though the field equations themselves may provide a fully satisfactory
account of all the basic phenomena known to date. From this point of view,
the ultimate origin of gravity  may lie in some fundamental degrees of
freedom quite different from the metric variables, namely degrees of
freedom which are completely unknown to us at
present\,\cite{Padmanabhan10}. If so, the field equations under discussion
in this paper could just be effective field equations falling in this
category and therefore no fundamental action to derive them would be
needed. A detailed discussion on this point goes beyond the scope of the
present work.

The quantities $\dot{H}$ and $H^2$ appearing on the \textit{r.h.s.} of
(\ref{entropica}) are related through $\dot{H}=-(q+1)\,H^2$ where $q$ is
the deceleration parameter. During some stages of the cosmic evolution
when $q$ is roughly constant, $\dot{H}$ and $H^2$ are approximately
proportional. For example, $q\simeq 1$ for the radiation dominated epoch,
and $q\simeq 1/2$ for the matter dominated epoch. Hence $\dot{H}\simeq
-2\,H^2$ deep in the radiation dominated era and $\dot{H}\simeq
-(3/2)\,H^2$ deep in the matter dominated epoch. When we compare the
entropic formula (\ref{entropica}) with the corresponding equation
(\ref{vacuuma}) in the running vacuum model, we see that deep in the
matter dominated epoch  we can set the correspondence $\nu\leftrightarrow
C_H-3\CHd/2$  between the two models. However this is not valid at low
redshifts when the universe goes over from matter domination to
accelerated expansion. In this interval where SNIa data are located, $q$
experiences a sharp model-dependent variation. Therefore the addition of
the term $C_{\dot H} {\dot H}$ is a genuine extension of the original
running vacuum energy model.

In view of the close analogy between these models, in the next section we
consider a generalization of the running vacuum model with the inclusion
of a term ${\dot H}$ together with the $H^2$ one.


\mysection{Background solution of the cosmological field equations}
\label{sect:solving}

In this section we consider the solution of the cosmological field
equations for both the generalized running vacuum model and the
entropic-force model. We discuss in detail the underlying local
conservation laws of matter and radiation in interaction with a dynamical
vacuum energy component and we show that this leads to important conceptual
issues. Finally, we emphasize the crucial importance of a
constant term which rules out some of the entropic-force models lacking
this term.

\mysubsection{The generic cosmological framework} \label{sect:solving1}

The cosmological equations of both the running vacuum models and the
entropic-force models can be solved in a common framework. We will
consider spatially flat Friedmann-Lema\^\i tre-Robertson-Walker (FRLW)
cosmologies
\be
ds^2 = dt^2 - a^{2}(t)~d{\bf x}^2~.\lb{RW}
\ee
Hence the (expansion) dynamics is fully encoded in the time evolution of
the scale factor $a(t)$. Instead of obeying the usual Friedmann equations
of General Relativity, our models obey modified Friedmann equations, viz. \bea
\biggl(\frac{\dot a}{a}\biggr)^2 &=& \frac{8\pi G}{3} ~ \sum_{i} \rho_i +
C_0 + C_H H^2 +
                                             C_{\dot H} {\dot H}\lb{FR1}\\
\frac{\ddot a}{a} &=& -\frac{4\pi G}{3} \sum_{i} (\rho_i + 3 p_i) + C_0 +
C_H H^2 +
                                             C_{\dot H} {\dot H} ~,\lb{FR2}
\eea
where the remaining sum is over the matter components only. For realistic
cosmologies, we take as usual two components, namely nonrelativistic
(dust-like) matter with $p_m = w_m~\rho_m = 0$ and radiation with $p_r =
w_r~\rho_r = \frac{1}{3}~\rho_r$. These equations can be viewed formally
as resulting from the presence of a time-dependent component
$\rho_{\Lambda}(t)=\rho_{\Lambda}(H(t),\dot{H}(t))$ satisfying \be
\rho_{\Lambda}(H,\dot{H}) =\frac{3}{8\pi G}\left( C_0 + C_H H^2 + \CHd
{\dot H}\right) = - p_{\Lambda}(H,\dot{H})~. \lb{GRVE} \ee
We will call this dynamical component a ``generalized running vacuum
energy'' (GRVE) density since its EoS satisfies
$w_{\Lambda}=p_{\CC}/\rho_{\CC}=-1$ as in the case of a strictly constant
vacuum energy\,\footnote{Let us notice that the recent
work\,\cite{Ademir12a}, which extends the discussion of a model first
suggested in Appendix C of \cite{BPS09}, contains a linear term in $H$
rather than our $\dot{H}$ term. As mentioned in
sect.\,\ref{sect:runningv},\ odd powers of $H$ cannot emerge from a
covariant effective action, and in this sense these models are more
phenomenological than the class of GRVE models (\ref{GRVE}) presented
here.} . By the same token we will call the class of these models with
$C_0\neq 0$ the ``generalized running vacuum models''. In the particular
case $C_{\dot H}=0$ we recover the original running vacuum model discussed
in sect.\,\ref{sect:runningv}. Formally, the generalization of the model
being proposed here implies that the scale $\mu^2$ in
Eq.\,(\ref{seriesLambda}), which is to be eventually associated with a
physical quantity according to the RG procedure, should in general be a
linear combination of $H^2$ and $\Hd$ rather than just the $H^2$
component, as these two terms represent independent d.o.f. with the same
dimension. Finally, let us emphasize that the particular case $C_0=0$ is
\textit{not} to be included within the class of GRVE models because the
integration of the RG equation (\ref{seriesLambda}) always involves an
additive term leading to $C_0\neq 0$. The case $C_0=0$ seems to appear in
some of the entropic-force models\,\cite{Easson10} briefly addressed in
the previous subsection. While this setting can be derived as a particular
case of our general analysis of the system (\ref{FR1})-(\ref{FR2}), we
stress that $C_0=0$ leads to a qualitatively new situation,  which we will
comment in subsequent sections and that is \textit{not} expected from the
conceptual point of view of the running vacuum model framework. With these
provisos in mind we are now going to solve the background cosmology of the
entire class of models (\ref{FR1})-(\ref{FR2}).

\mysubsection{Discussion of the local conservation laws}
\label{sect:solving2}

Once the metric (\ref{RW}) is given, a comoving perfect fluid with
energy-momentum tensor $T_{\mu\nu} = (\rho_i + p_i)~u_{\mu} u_{\nu} -
p_i~g_{\mu\nu}$ will satisfy the conservation equation ${\dot \rho_i} = -
3 H (\rho_i + p_i)$, with $\rho_i$ and $p_i$ appearing in
(\ref{FR1})-(\ref{FR2}) provided $\rL$ is constant. This applies of course
for non-flat FLRW universes as well. However, if the vacuum energy density
is a time-dependent component it cannot have this energy-momentum tensor.
Actually, as we will see now the same applies for the other components
appearing in equations (\ref{FR1})-(\ref{FR2}).

However, from the system (\ref{FR1})-(\ref{FR2}) we get the coupled
conservation equation \be {\dot \rho}_m + {\dot \rho}_r + {\dot
\rho}_{\Lambda} = -3 H \rho_m - 4 H \rho_r \lb{ceq1}\,. \ee This equation
is indeed a first integral of that system. Clearly, for ${\dot
\rho}_{\Lambda}\ne 0$ none of the components can satisfy the standard
conservation equation as emphasized above\,\footnote{The non-conservation
of matter in the presence of running vacuum energy has recently been
proposed as a possible link between the dynamical DE and the increasing
evidence for a possible variation of the fundamental constants and scales
in Nature, as e.g. the QCD scale -- see \cite{FrizschSola12}. However, a
running vacuum energy of the form (\ref{RGlaw2}) can be made compatible
with matter conservation if one allows $G$ to slowly evolve with time, see
\cite{Fossil07} for a concrete scenario connected with an action
functional. For the present GRVE framework, though, $G$ is assumed to be
strictly constant.}.

It is easy to see that this does not depend on the particular choice
(\ref{GRVE}). Let us assume that the variable vacuum has an energy density
$\frac{8\pi G}{3}~\rho_{\Lambda}(t)= C_H~H^2$ {\emph{and}} that it is a
perfect fluid with an a priori undefined equation of state. Then it
follows from its (assumed) conservation equation that its equation of
state parameter $w_{\Lambda}$ satisfies $-\frac{\dot
H}{H^2}=\frac{3}{2}(1+w_{\Lambda})$. So if we consider a simple (flat)
universe containing also dust, the only consistent way is to have
$\Omega_{\Lambda}=C_H=1$. So we end with no dust at all and the first
Friedmann equation reduces to an equality (while $w_{\Lambda}$ remains of
course undefined)! We stress that these conceptual issues are also true
for other models with similar effective Friedmann equations, for example
models inspired by the holographic principle or models based on the
entropic-force principle. Ultimately these properties arise from the
absence (in general) of a formulation at the level of the action which is
still an open issue.

As mentioned in the previous section, in a pragmatic approach we assume
the validity of (\ref{FR1})-(\ref{FR2}) without explicitly deducing them
from an underlying action -- see, however, \cite{Fossil07} for a specific
framework along these lines. After some calculations the coupled
conservation equation (\ref{ceq1}) reads \be {\dot \rho}_m + {\dot \rho}_r
- \frac{3}{2} C_{\dot H} \left( {\dot \rho}_m + \frac{4}{3}{\dot \rho}_r
\right) = -3 H (1 - C_H) \left( \rho_m + \frac{4}{3} \rho_r
\right)~.\lb{ceq} \ee We note that the model does not yield a conservation
equation for each component separately, seemingly overlooked
in\,\cite{Easson10}. For this we would need to specify the action of the
model and to find the corresponding energy-momentum tensors. Moreover, the
system is not fully defined by eqs.(\ref{FR1})-(\ref{FR2}). Indeed, any
solution of the following set of equations \bea
{\dot \rho}_m &=& -3 H \frac{ 1 - C_H }{ 1 - \frac{3}{2} C_{\dot H} }~\rho_m  + Q \lb{dm1} \\
{\dot \rho}_r &=& -4 H \frac{ 1 - C_H }{ 1 - 2 C_{\dot H} }~\rho_r - Q~,\lb{dr1}
\eea
for arbitrary function $Q=Q(t)$, will be a solution
of equation (\ref{ceq}). In the matter dominated ($\rho_r\approx 0$) and
radiation dominated ($\rho_m\approx 0$) stages we must have from
(\ref{ceq}) that $Q\to 0$. The simplest version of this model is to assume
that eq.(\ref{ceq}) reduces at all times to a set of decoupled equations
with $Q=0$ at all times. We conjecture that this is perhaps the only way
to introduce consistently an arbitrary number of species.

It will be convenient to introduce the following notations
\bea
\nu & \equiv & C_H\lb{nu}\\
\alpha & \equiv & \frac{3}{2} C_{\dot H}~,\lb{al}
\eea
as well as the important quantities
\bea
\xi_m & \equiv & \frac{ 1 - \nu }{ 1 - \alpha } \lb{xim}\\
\xi_r & \equiv & \frac{ 1 - \nu }{ 1 - \frac{4}{3}\alpha }~.\lb{xir}
\eea
The motivation for the relabeling (\ref{nu}) is simply because for
$\CHd=0$ the GRVE model boils down to the original running vacuum model
discussed in sect.\,\ref{sect:runningv}, and then $C_H$ exactly reduces to
the parameter $\nu$ defined in that section. Using these definitions,
equations (\ref{dm1}) and (\ref{dr1}) (setting $Q=0$) can be written as
follows \bea
{\dot \rho}_m &=&  -3 H ~\xi_m ~\rho_m \lb{dm} \\
{\dot \rho}_r &=&  -4 H \xi_r ~\rho_r ~,\lb{dr}
\eea
for which it is straightforward to obtain the corresponding solutions (setting $a_0=1$):
\bea
%
\rho_m &=& \rho_m^0 ~a^{-3 \xi_m}
                   =  \rho_m^0~(1 + z)^{3 \xi_m}     \lb{mz} \\
\rho_r &=& \rho_r^0 ~a^{-4 \xi_r}
                   = \rho_r^0~(1 + z)^{4 \xi_r}~.\lb{rz}
\eea
Note that these decoupled solutions reduce automatically to the
behavior of dustlike matter during matter domination ($\rho_r\approx 0$)
and to the radiation component during radiation domination ($\rho_m\approx
0$). They take the standard form for $\xi_m=1$ and $\xi_r=1$.

\mysubsection{Determining the time evolving vacuum energy and the Hubble
function} \label{sect:solving3}

The equations (\ref{dm}) and (\ref{dr}) are decoupled, there is no
transfer of energy between the two components. However there is a transfer
of energy between the running vacuum energy $\rho_\Lambda{}$ and these
components. We find for the evolution of $\rho_{\Lambda}$ \be {\dot
\rho}_{\Lambda} = 3 H (\xi_m -1)~\rho_m + 4 H (\xi_r -1)~\rho_r~. \lb{dL}
\ee
When $\xi_m = \xi_r = 1$  the standard behavior of matter and radiation is
recovered and then $\rho_{\Lambda}=\CC/(8\pi\,G)$ reduces to a genuine
cosmological constant with $\Lambda=3~C_0$. The transfer of energy between
the matter components and the GRVE is the physical reason for the
particular scaling behaviors (\ref{mz}) and (\ref{rz}), which obviously
depart from the standard expectations $\rho_m\sim a^{-3}$ and $\rho_r\sim
a^{-4}$ in the $\CC$CDM owing to the non-vanishing values of the
parameters $\nu$ and $\alpha$. We have here an effective interacting dark
energy model. Consistency enforces to have an interaction between the
running vacuum energy and {\emph{all}} other components.

The addition of the term $C_{\dot H}~{\dot H}=\frac{2}{3}\alpha~{\dot H}$
introduces an important change compared to the original running vacuum
model discussed in sect.\,\ref{sect:runningv} -- in which $C_{\dot H}=0$
-- but the new degree of freedom is severely constrained by observations.
The reason is that the model cannot depart too much from the $\CC$CDM
values $\xi_m = \xi_r = 1$, with \be  \xi_r = \xi_m ~\frac{ 1 - \alpha }
{ 1 - \frac{4}{3}\alpha }~. \ee The only way to satisfy {\emph{both}}
constraints $\xi_m\approx 1$ and $\xi_r\approx 1$ is to have
\be |\nu|\ll 1\,,\ \   |\alpha|\ll 1\,\,\,\iff \,\,\, \xi_m\approx 1\,,\ \
\xi_r\approx 1 \lb{null1} \ee Of course we have $\xi_m =
\xi_r = 1$ when both parameters $\nu$ and $\alpha$ vanish. Note also that
the condition $\xi_r\approx 1$ is crucial for the viability of our model
at high redshifts (e.g. when fitting the model against CMB data). If we
use only constraints at very low redshifts, models with $\xi_m\approx 1$
(i.e. $|\nu|\ll 1$) but not necessarily satisfying $|\alpha|\ll 1$ will
fare well in this domain, though the model would actually be unviable
taking into account its behavior at high redshifts.

Compared to the old running vacuum energy model ($\alpha = 0$), the
generalized model offers more possibilities to depart from standard
cosmology:
\begin{itemize}
\item One can have $\xi_r = \xi_m\approx 1$, in which case the model
    just reduces to the original running model ($\alpha = 0$ or
    $C_{\dot H}=0$, and $\nu\ne 0$). Both radiation and dust scale in
    a non-standard way but their departure from standard behavior is
    not independent and depends on one single parameter $\nu$,
    specifically: $\rho_m\sim a^{-3(1-\nu)}$ and $\rho_r\sim
    a^{-4(1-\nu)}$.

\item $\xi_r = 1$ and $\xi_m\approx 1$, in which case radiation
    behaves in the standard way but dust does not. This occurs when
    $\nu = \frac{4}{3} \alpha\ne 0$. This case exists only in the
    generalized model $\alpha\ne 0$. Departure from standard cosmology
    occurs already at low redshifts.

\item $\xi_m = 1$ and $\xi_r\approx 1$, now dust scales in standard
    way but radiation does not. This corresponds to $\nu = \alpha\ne
    0$. This case can mimic standard cosmology at low redshifts but it
    is strongly constrained when high redshift data are considered;

\item $\xi_m\approx 1$, $\xi_r\approx 1$ and $\xi_m\ne \xi_r$. Here the
    deviation of the non-relativistic component is different from the
    relativistic one, and hence this provides an extension of the
    first case discussed above which is only possible within the GRVE
    model.

\end{itemize}

The three last cases above are only possible in the generalized model
$\alpha\ne 0$. However these additional possibilities are strongly
constrained by observations on both low and high redshifts. We will see in
particular in sect.\,\ref{sect:fitting} that the strong constraint on
$\alpha$ in the regime (\ref{null1}) is similar to that one found for
$\nu$ in the original running model -- see the recent
analyses\,\cite{BPS09,GSBP11}.

Equation (\ref{dL}) is easily recast in the form \be \frac{d
\rho_{\Lambda}}{da} = 3~(\xi_m -1)~\frac{\rho_m}{a} +  4~(\xi_r
-1)~\frac{\rho_r}{a}\lb{dLa} \ee which is easily integrated using the
solutions (\ref{mz}) and (\ref{rz}) and the explicit form of the vacuum
energy as a function of the scale factor can be expressed as follows:
\begin{equation}\label{eq:rLa}
\rL(a)=\rLo+{\rMo}\,\,(\xiM^{-1} - 1) \left( a^{-3\xiM} -1  \right) +
{\rRo}\,\,(\xiR^{-1} - 1) \left( a^{-4\xiR} -1\right)\,.
\end{equation}
The Hubble function can now be constructed from the matter components
(\ref{dm})-(\ref{dr}) and the vacuum energy (\ref{eq:rLa}):
\begin{equation}\label{eq:Hubblerate} H^2=\frac{8\pi\,G}{3}\left[\rM+\rR+\rL\right]=
H_0^2\left[\OMo~\frac{\rho_m}{\rMo} + \ORo~\frac{\rR}{\rRo}+
\OLo~\frac{\rL}{\rLo}\right]\,.\end{equation}
Introducing the normalized Hubble rate in terms of the redshift,
$E(z)\equiv {H(z)}/{H_0}$, we find: \be E^2(z) =
\frac{\OMo}{\xi_m}~(1+z)^{3 \xi_m} + \frac{\ORo}{\xi_r}~(1+z)^{4 \xi_r}
                           + \frac{ H_0^{-2}C_0 }{ 1 - \nu }~,\lb{E2z}
\ee where we have used the standard definition $\Omega_i =
{\rho_i}/{\rho_{c}}$, with $\rc=3H^2/(8\pi G)$, satisfying the constraint
\be \Omega_{\Lambda} + \Omega_m + \Omega_r = 1~, \ee at all times.

Note that the boundary condition $E(z=0)=1$ in (\ref{E2z}) leads to the
equality \be H_0^{-2}C_0 = \OLo - \Delta\nu~,\lb{Delnu} \ee where we have
defined $\Delta\nu=\nu -\bar{\nu}$, with $\bar{\nu}=\alpha\,\OMo+
(4/3)\,\alpha\,\ORo$. The $\bar{\nu}$ parameter is characteristic of the
extension of the original running vacuum model into the GRVE model and is
closely related to  $\nu$. Indeed $\bar{\nu}$ gauges the size of the new
$\dot{H}$-effect in terms of $H^2$ at the present time since it satisfies
the relation $\CHd\,\dot{H}_0=-\bar{\nu}\,H_0^2$, which can be compared to
$C_H\,H_0^2=\nu\,H_0^2$ in the original running model. To confirm that relation let
us write the current value of the $\CHd\,\dot{H}$ term in the starting
equations (\ref{FR1})-(\ref{FR2}) as follows
\begin{equation}\label{eq:defbarnu}
\CHd\,\dot{H}_0=-\CHd\,(q_0+1)\,H_0^2=-\left(\frac32\,\OMo+2\,\ORo\right)\,\CHd\,H_0^2
\,.
\end{equation}
Thus we find $\bar{\nu}=\bar{\nu}_m+\bar{\nu}_r$, where
$\bar{\nu}_m=(3/2)\,\OMo\,\CHd=\alpha\,\OMo$ and
$\bar{\nu}_r=2\,\ORo\,\CHd=(4/3)\,\alpha\,\ORo$ represent the
non-relativistic and relativistic matter contributions respectively.

After having determined the explicit relation between $\bar{\nu}$ and the
other parameters, we see from  Eq.\,(\ref{Delnu}) that $C_0$ becomes also
explicitly determined as follows:
\begin{equation}\label{eq:Coexplicit}
C_0 =H_0^{2}\left[\OLo-\nu+\,\left(\OMo+\frac43\,\ORo\right)\,\alpha\right]=\,H_0^{2}\left[\OLo-C_H+\frac32\left(\,\OMo+\frac43\,\ORo\right)\,\CHd\right]\,.
\end{equation}
Notice that for $\alpha=0$ (or $\CHd=0$) it boils down to the
corresponding expression (\ref{C0}) for the original running vacuum model.
On the other hand Eq.\,(\ref{eq:Coexplicit}) tells us another interesting
feature, to wit: models with $C_0=0$ cannot have the two parameters $\nu$
and $\alpha$ (equivalently $C_H$ and $\CHd$) simultaneously small, i.e. it
is impossible to satisfy the relations (\ref{null1}), unless $\OLo=0$ --
which is of course unacceptable. In particular, entropic-force
models\,\cite{Easson10} cannot have $C_H$ and $\CHd$ simultaneously small,
otherwise they would contradict the measured value of the cosmological
term: $\OLo\simeq 0.73$. Even if we would accept that at least one of the
parameters $C_H$ and $\CHd$ is not small, the resulting model would be
contrived as it would entail a non-trivial modification of the standard
$\CC$CDM cosmology. Actually in the next section we will encounter a
related difficulty, which is perhaps the biggest stumbling block to the
$C_0=0$ models.

\mysubsection{Crucial distinction between some entropic-force and GRVE
models}\label{sect:solving4}

In the previous subsections we have solved in detailed the full class of
cosmological models based on the set of generalized FLRW equations
(\ref{FR1})-(\ref{FR2}). In particular, we have assumed
arbitrary values for the parameters $C_0$, $C_{H}\equiv\nu$ and
$\CHd\propto\alpha$. However we expect from observational constraints
that the last two ones are sufficiently small -- cf.  Eq.\,(\ref{null1})
-- in order for the generalized models not to depart too much from the
standard scaling laws of matter and radiation.

We turn now our attention specifically to the additive parameter $C_0$. If
the other two parameters ($C_H$ and $\CHd$) have to be small, this is not
the case for $C_0$ and as we will see now it cannot vanish. While the
running vacuum energy models have a nonvanishing $C_0$, this is not the
case for some entropic force models.

Indeed, successful models must be able to produce an accelerated expansion
at very low redshifts. To start with let us analyze the situation $C_0=0$.
It is easy to derive from the expression for $\frac{\ddot a}{a}$ that
accelerated expansion is obtained both in the matter and
radiation-dominated stages if the following condition is satisfied (with
$C_0=0$) \be \nu - \frac{2}{3}\alpha >
\frac{1}{2}~~~~~~~~~~~~{\iff}~~~~~~~~~~~~~2~\xi_r < 1~.\lb{stacc} \ee In
the matter-dominated era a slightly weaker condition is required \be \nu -
\frac{2}{3}\alpha >
\frac{1}{3}~~~~~~~~~~~~{\iff}~~~~~~~~~~~~~\frac{3}{2}~\xi_m < 1~.\lb{wacc}
\ee We see that these conditions are redshift independent. Therefore, if
we have an accelerated expansion rate at very low redshifts, we will have
it at least during all of the matter-dominated stage. This leads obviously
to an unviable cosmology putting aside the fact that the corresponding
scaling behaviors are completely unviable observationally.

We can recover these results solving for the time dependence of $H(t)$ and $a(t)$. The following
equation holds during matter domination
\be
{\dot H} + \frac{3}{2}~\xi_m~H^2 = \frac{3}{2}~\xi_m~\frac{C_0}{1 - \nu}\lb{dHm}
\ee
When $C_0=0$ the condition (\ref{wacc}) for accelerated expansion is clearly recovered from (\ref{dHm}).
In the general case  $C_0\ne 0$, equation (\ref{dHm}) can be solved to yield
\bea
H(t) &=& A~\coth \left[ \frac{3}{2}~\xi_m~A~t \right]~,\\
a(t) &=& D ~\sinh ^{\frac{2}{3\xi_m} }
            \left[ \frac{3}{2}~\xi_m~A~t \right]~.
\eea
where we have used (\ref{Delnu}) and we have set
\be
D =  \left[ \frac{ \Omega_{m,0} - \bnu_M )}{ \Omega_{\Lambda,0} - \Delta\nu } \right]^{\frac{1}{3\xi_m}}
~~~~~~~~~~~~~~~A = H_0~\sqrt{ \frac{\Omega_{\Lambda,0} - \Delta\nu}{1-\nu}}~.
\ee
Returning to the case with vanishing $C_0$, the solution to (\ref{dHm}) reads
\be
a(t) \propto t^{\frac{2}{3\xi_m}}~~~~~~~~~~~~~~~~~~~~\frac{\ddot{a}}{a}\sim
\left(\frac{2}{3\xiM}-1\right)\,t^{-2}~,
\ee
which shows again that (\ref{wacc}) leads to accelerated expansion.
Hence we conclude that models with $C_0=0$ cannot describe an expanding universe
undergoing a transition from decelerated to accelerated expansion.

\mysubsection{Observational interpretation}

For an ``Einsteinian'' interpretation of the generalized running vacuum
models, i.e. those represented by equations
(\ref{eq:Hubblerate})-(\ref{E2z}) with $C_0\neq 0$, let us write \bea H^2
= \frac{8\pi G}{3}\left[\tilde{\rho}_m + \tilde{\rho}_r\right] +
\frac{\tilde{\Lambda}}{3} = H_0^2\,\left[
\tilde{\Omega}_{m}^{\,0}~\frac{\tilde{\rho}_m}{\tilde{\rho}_{m}^{\,0}} +
        \tilde{\Omega}_{r}^{\,0}~\frac{\tilde{\rho}_r}{\tilde{\rho}_{r}^{\,0}}\right] +
        \frac{\tilde{\Lambda}}{3}\,,\lb{Einsteinian}
\eea with the obvious identifications \be \tilde{\rho}_m =
\frac{\rho_m}{\xi_m}\,,~~~~~~\tilde{\rho}_r =\frac{\rho_r}{\xi_r}\,,~~~~~~
\frac{\tilde{\Lambda}}{3} = \frac{C_0}{1 - \nu}\,,~~~~~~
\tilde{\Omega}_{m}^{\,0}
=\frac{\OMo}{\xi_m}=\frac{\tilde{\rho}_m^0}{\rco}\,,~~~~~~\tilde{\Omega}_{r}^{\,0}
=\frac{\OMo}{\xi_r}=\frac{\tilde{\rho}^0_r}{\rco}~.\lb{H2RG} \ee
Observationally there is no reason to distinguish between the matter or
radiation energy density appearing in the starting equation (\ref{FR1})
and that part contained in $\rho_{\Lambda}$. Hence it is natural to
 identify the observed value $\Omega_{m, {\rm obs}}^0$ with
$\tilde{\Omega}_{m}^{\,0}$ and similarly $\Omega_{r;{\rm obs}}^0$ with
$\tilde{\Omega}_{r}^{\,0}$. We still have the standard equality valid at
all times \be \tilde{\Omega}_{m} + \tilde{\Omega}_{r} +
\tilde{\Omega}_{\Lambda} =1~, \lb{SOm} \ee with
$\tilde{\Omega}_{\Lambda} = \frac{\tilde{\Lambda}}{3 H^2}$.

Even recast in the form (\ref{Einsteinian}) we should remember that the
energy densities $\tilde{\rho}_m$ and
$\tilde{\rho}_r$, obey the nonstandard scaling laws (\ref{mz}), resp.
(\ref{rz}). Interestingly, in this ``Einsteinian'' interpretation, our
model reduces to a model with a genuine cosmological constant
$\tilde{\Lambda}$ and nonstandard evolution of dust and radiation. In the
generalized running vacuum energy model, the departure from standard
behavior of dust and radiation are independent from each other.

In this model, the redshift at equality $z_{eq}$ is given by \be \left( 1
+ z_{eq} \right)^{4\xi_r-3\xi_m} = \frac{ \tilde{\Omega}_{m}^{\,0} }{
\tilde{\Omega}_{r}^{\,0} }\lb{zeq} \ee The variation of $z_{eq}$
constrains the quantity $4\xi_r-3\xi_m$, a constraint that will be
satisfied by our best-fit models found in next section. In view of
(\ref{E2z}) we expect further very tight constraints on $\xi_r$ itself
deep in the radiation dominated era. 

In a first conservative approach we would like to keep a standard 
thermal history. Even if $\tilde{\Omega}_{r}^{\,0}$ assumes the value 
for $\Omega_{r}^{\,0}$ required by standard Big Bang Nucleosynthesis 
(BBN), and assuming that cosmic temperature scales in the standard way, 
the expansion rate at the BBN epoch will get changed by a non-standard 
amount due to the scaling law (\ref{rz}). Inserting numbers this 
finally yields the conservative constraint $|\xi_r - 1|<10^{-3}$ 
because the expansion rate is severely constrained and cannot vary too 
much at the time of BBN (see e.g. \cite{J09}).

As $\xi_r$ is a free parameter, in practice we wish to explore scenarios 
satisfying
\be 
\xi_r = 1 \lb{xir1}\,. 
\ee
This choice means that the model parameters $\nu$ and $\alpha$ are no
longer independent, and from (\ref{xir}) we see that we must have
$\alpha=3\nu/4$. This ensures that the standard thermal history is 
recovered. Indeed, with the choice (\ref{xir1}), the radiation dominated
stage in our models is essentially similar to the standard radiation
dominated stage. We have in particular that the temperature of
thermalized relativistic species scales consistently in the standard way. 
This is in particular true for the Cosmic Microwave Background (CMB) 
temperature.

We have derived all equations necessary in order to constrain with
observations the class of generalized FLRW models (\ref{FR1})-(\ref{FR2})
satisfying (\ref{xir1}).  This we do in the next section.


\mysection{Fitting the models to the observational data}
\label{sect:fitting}

In the following we present some details of the statistical method and on
the observational samples and data statistical analysis that will be
adopted to constrain the models presented in the previous sections. We
shall extract our fit from the combined data on type Ia supernovae (SNIa),
the data on the Baryonic Acoustic Oscillations (BAOs), and the shift
parameter of the Cosmic Microwave Background (CMB). Note that in the case
of the BAO analysis we have to modify it appropriately in order to
incorporate some specific features of the present models.

\mysubsection{The global fit to SNIa, BAOs and CMB}

First of all, we use the {\em Union 2} set of 557 type Ia supernovae of
Amanullah et al.\,\cite{Ama11}\,\footnote{Note that the data can be found
in: http://supernova.lbl.gov/Union/.}. The corresponding
$\chi^{2}$-function to be minimized is:
\begin{equation}\label{eq:xi2SNIa}
\chi^{2}_{\rm SNIa}({\bf p})=\sum_{i=1}^{557} \left[ \frac{ {\cal
\mu}_{\rm th} (z_{i},{\bf p})-{\cal \mu}_{\rm obs}(z_{i}) }
{\sigma_{i}} \right]^{2} \;,
\end{equation}
where $z_{i}$ is the observed redshift for each data point. The fitted
quantity ${\cal \mu}$ is the distance modulus, defined as ${\cal
\mu}\equiv m-M=5\log{d_{L}}+25$, in which $d_{L}(z,{\bf p})$ is the
luminosity distance:
\begin{equation}\label{eq:LumDist}
d_{L}(z,{\bf p})={(1+z)} \int_{0}^{z} \frac{{\rm d}z'}{H(z')}\;.
\end{equation}
Here  ${\bf p}$ a vector containing the cosmological parameters of our
model that we wish to fit for. In our case one possibility would be to
take e.g. ${\bf p}=(\tilde{\Omega}_{m}^{\,0},\nu)$. In equation
(\ref{eq:xi2SNIa}), the theoretically calculated distance modulus
$\mu_{\rm th}$ for each point follows from (\ref{eq:LumDist}), in which
the Hubble function $H(z)=H_0\,E(z)$ is given by (\ref{E2z}) for the
generic model under consideration. Finally, $\mu_{\rm obs}(z_{i})$ and
$\sigma_i$ stand for the measured distance modulus and the corresponding
$1\sigma$ uncertainty for each SNIa data point, respectively. The previous
formula (\ref{eq:LumDist}) for the luminosity distance applies only for
spatially flat universes, which we are assuming throughout. {Note that
since only the relative distances of the SNIa are accurate and not their
absolute local calibration, we always marginalize with respect to the
internally derived Hubble constant (for methods that do not need to a
priori marginalize over the internally estimated Hubble constant, see for
example \cite{NesPer,Wei08}). In the case of the {\it Union2} SNIa data
the internally derived Hubble constant is $H_0\simeq 70$Km/s/Mpc which is
in agreement to that of WMAP7 \cite{WMAP} $H_0=70.4$Km/s/Mpc used in the
present study.}

In addition to the SNIa data, we also consider the BAO scale produced in
the last scattering surface by the competition between the pressure of the
coupled baryon-photon fluid and gravity. The resulting acoustic waves
leave (in the course of the evolution) an overdensity signature at certain
length scales of the matter distribution. Evidence of this excess has been
found in the clustering properties of the SDSS galaxies
(see~\cite{Eis05},~\cite{Perc10,Blake11}) and it provides a ``standard
ruler'' that we can employ to constrain dark energy models. In this work
we use the results of Percival et al. \cite{Perc10}, $r_{s}(z_{d})/D_{\rm
V}(z_{\star})=0.1390\pm 0.0037$. Note that $r_{s}(z_d)$ is the comoving
sound horizon size at the baryon drag epoch\,\cite{Eis98} (i.e. the epoch
at which baryons are released from the Compton drag of photons), and
$z_{d}\sim {\cal O}(10^{3})$ is the corresponding redshift of that epoch,
closely related to that of last scattering-- the precise expression being
given by the fitting formula of \cite{Eis98}. Finally, $D_{V}(z)$ is the
effective distance measure \cite{Eis05} and $z_{\star}=0.275$. Of course,
the quantities $(r_{s},D_{\rm V})$ can be defined analytically. In
particular, $r_{s}(z_d)$ is given by the comoving distance that light can
travel prior to redshift $z_d$:
\begin{equation}
\label{drag}
r_{s}(z_{d})=\int_{0}^{t(z_d)}\,\frac{c_s\,dt}{a}=\int_{0}^{a_{d}}
\frac{c_s(a)\,da}{a^{2} H(a)}\;,
\end{equation}
where $a_{d}=(1+z_{d})^{-1}$, and
\begin{equation}\label{eq:cs2}
c_s(a)=\left(\frac{\delta \tilde{p}_{\gamma}}{\delta \tilde{\rho}_{\gamma}+\delta \tilde{\rho}_b}\right)^{1/2}=
\frac{1}{\sqrt{3\,\left(1+{\cal R}(a)\right)}}
\end{equation}
is the sound speed in the baryon-photon plasma. Here we assume adiabatic perturbations and we
have used $\delta \tilde{p}_b=0$ and $\delta \tilde{p}_{\gamma}=(1/3)\,\delta\tilde{\rho}_{\gamma}$,
and defined
${\cal R}(a)=\delta\tilde{\rho}_b/\delta\tilde{\rho}_{\gamma}$. If the scaling
laws for non-relativistic matter and radiation would be those of the standard model, we
would have
${\cal R}(a)=3\rho_b/4\,\rho_{\gamma}$, which
can be finally cast as a linear function of the scale factor:
${\cal R}(a)=\left({3\Omega_{b}^{0}}/{4\Omega_{\gamma}^0}\right)\,a$, where
$\Omega_{b}^{0}h^{2}\simeq 0.02263$ and $\Omega_{\gamma}^0\,h^2\simeq
2.47\times 10^{-5}$ are the current values of the normalized baryon and
photon densities. However, our scaling laws for non-relativistic matter and radiation are
given by equations (\ref{mz}) and (\ref{rz}). As a result, the sound speed
velocity in the plasma gets a correction with respect to the standard result ($a_0=1$):
\begin{equation}\label{eq:Ra2}
{\cal R}(a)=\frac34\,\frac{\xiM}{\xiR}\,\frac{\tilde{\rho}_b(a)}{\,\tilde{\rho}_{\gamma}(a)}
=\frac34\,\frac{1-4\alpha/3}{1-\alpha}\frac{\tilde{\Omega}_{b}^{0}}{\tilde{\Omega}_{\gamma}^0}\,
a^{4\xiR-3\xiM}\,.
\end{equation}
Of course for $\nu=0$ and $\alpha=0$ ($\xiM=\xiR=1$) the
previous equation becomes again a linear function of the scale factor,
and it exactly reduces to the standard result.

The remaining ingredients of the BAO analysis are as in the standard case,
in particular the effective distance is (see \cite{Eis05}):
\begin{equation}
D_{\rm V}(z)\equiv \left[ (1+z)^{2} D_{A}^{2}(z) \frac{cz}{H(z)}\right]^{1/3}\;,
\end{equation}
where $D_{A}(z)=(1+z)^{-2} d_{L}(z,{\bf p}) $ is the angular diameter
distance. Therefore, the corresponding $\chi^{2}_{\rm BAO}$ function is
simply written as:
\begin{equation}
\chi^{2}_{\rm BAO}({\bf p})=
\frac{\left[\frac{r_{s}(z_{d})}{D_{\rm V}(z_{\star})}({\bf
p})-0.1390\right]^{2}}{0.0037^{2}} \;.
\end{equation}

Furthermore, a very accurate and deep geometrical probe of dark energy is
the angular scale of the sound horizon at the last scattering surface, as
encoded in the location $l_1^{TT}$ of the first peak of the Cosmic
Microwave Background (CMB) temperature perturbation spectrum. This probe
is described by the  CMB shift
parameter~\cite{Bond:1997wr,Nesseris:2006er}, defined as:
\begin{equation}\label{eq:shiftparameter}
R=\sqrt{\OMo}\int_{0}^{z_{ls}} \frac{dz}{E(z)}\,.
\end{equation}
The measured shift parameter according to the WMAP 7-years
data~\cite{WMAP} is $R=1.726\pm 0.018$ at the redshift of the last
scattering surface: $z_{ls}=1091.36$. In this case, the
$\chi^{2}$-function is given by:
\begin{equation}
\chi^{2}_{\rm CMB}({\bf p})=\frac{[R({\bf
p})-1.726]^{2}}{0.018^{2}}\;.
\end{equation}
For a detailed discussion of the shift parameter as a cosmological probe,
see e.g.\,\cite{Elgaroy07}. Let us emphasize that when dealing with the
CMB shift parameter we have to include both the matter and radiation terms
in the total normalized matter density entering the $E(z)$ function in
(\ref{eq:shiftparameter}), given explicitly by (\ref{E2z}). Indeed, the
radiation contribution reads $\ORo=(1+0.227 N_{\nu})\, \Omega_{\gamma}^0$,
with $N_{\nu}$ the number of neutrino species.
Therefore, at $z_{ls}=1091.36$, and including three light neutrino
species, the radiation contribution amounts to $\sim 24\%$ of the total
energy density associated to matter, which is not negligible. We use
$h=0.704$ in our analysis.

Our statistical analysis, due to its simplicity, has been used
extensively in the literature in order to constrain the dark energy models
(see for example \cite{essence,Wei08,Pli11} and references therein).
We would like to point that a more general statistical
presentation would require the covariances of BAO and CMB shift parameter.
We have checked our statistical results using the latter covariances and
our results remain the same as they should. Note that the corresponding
covariances can be found in Percival et al. \cite{Perc10} and in
Komatsu et al. \cite{WMAP} respectively.
{Finally, as emphasized before eq.(3.1.1) we restrict our analysis to
spatially flat spaces. This seems justified in view of the tight
constraints on $\Omega_{k,0}$ and is sufficient for our purposes.}

%

\mysubsection{Numerical results}

Since we perform an overall fit of the SNIa+BAO+CMB data, it is important
to take into account the contribution of both non-relativistic matter and
radiation.

\begin{itemize}
\item For the concordance $\Lambda$CDM cosmology, we simply have
    $\rLo=$const. and
\begin{equation}\label{mixture}
\rmr(z)=\rmo\,(1+z)^{3}\,, \ \ \ \ \ \ \ \ \rR(z)=\rRo\,(1+z)^{4}\,.
\end{equation}

\item Concerning the generalized running vacuum energy model
    (\ref{FR1})-(\ref{FR1}) we have explicitly given the corresponding
    density formulae in sect.\,\ref{sect:solving}. Let us recall that
    $C_0=0$ is a very particular case that we exclude from the class
    of the GRVE models. We have shown in the previous section that
    this case is not viable observationally and therefore we will not
    consider it any further for the phenomenological analysis.
    Therefore, from now on we assume that $C_0\neq 0$ and focus on
    fitting the parameters of this model to the SNIa+BAO+CMB data. In
    particular, we already know that $\nu$ and $\alpha$ have to be
    small -- see Eq.\,(\ref{null1}) -- but only the direct
    confrontation of the model with the data will tell us about their
    possible maximum size. In practice, considering models satisfying
    (\ref{xir1}), it will be convenient to define the effective
    parameter
    \begin{equation}\label{eq:nueff}
    \nueff\equiv \nu-\alpha = \frac{1}{4}\nu \,,
    \end{equation}
    and use $\nueff$ as fitting parameter, together with
    $\tilde{\Omega}_m^0$. We can check this explicitly by expanding $\xiM$ and
    $\xiR$ linearly in $|\nu|\ll 1$ and $|\alpha|\ll 1$, together with
    some coefficients in (\ref{eq:rLa}), and using the definition
    (\ref{eq:nueff}):
\begin{equation}\label{eq:xiMxiRlinear}
    \xiM\simeq 1-\nueff\,,\ \ \ \ \ \ \xiR=1\,.
    \end{equation}
As a result, for the energy densities we find:
\begin{equation}\label{mixturev}
\tilde{\rho}_m(z) = \tilde{\rho}_m^0(z)\,(1+z)^{3(1-\nueff)}\,,\ \ \ \ \ \
                           \tilde{\rho}_r(z)= \tilde{\rho}_r^0\,(1+z)^{4}\,.
\end{equation}
Similarly the corresponding normalized Hubble flow squared reads
\begin{equation} \label{HubbleFlow3}
E^2(z) = \tilde{\Omega}_m^0\,(1+z)^{3(1-\nueff)} +
\tilde{\Omega}_r^0\,(1+z)^4 + \tilde{\Omega}^0_{\Lambda}~.
\end{equation}
Finally, let us mention that within the same approximation we can write
the BAO ratio (\ref{eq:Ra2}) entering the modified sound speed of the
baryon-photon plasma as follows:
\begin{equation}\label{eq:Ra3}
{\cal R}(a)=\frac34\,(1-\nueff)~\frac{\tilde{\Omega}_{b}^{0}}{\tilde{\Omega}_{\gamma}^0}\,
a^{1+3\nueff}\,.
\end{equation}
For $\nu=0$ and $\alpha=0$ (hence $\nueff=0$) it clearly reduces to
the standard result mentioned in the previous subsection.

These formulae confirm our contention that we can fully reexpress all
    the background formulae in terms of the effective fitting vector
   \begin{equation}\label{eq:peff}
    {\bf p}_{\rm eff}=\left(\tilde{\Omega}_{m}^{0},\nueff\right)\,.
    \end{equation}
    We also see from the previous formulae that $\nueff$ is the single
    effective parameter that controls the deviations of the GRVE model
    with respect to the $\CC$CDM model in the low $z$ region (when
    radiation can be neglected). It is only in the high redshift
    region where the model is sensitive to independent contributions
    from $\nueff$ (equivalently, from $\nu$ or $\alpha$). Notice that
    this feature could be used, in principle,
    to distinguish between the two sorts of running models, i.e. the original
    one (which we reviewed briefly in sect.\,\ref{sect:runningv}) and
    the generalized running vacuum model under discussion in this
    paper. At low $z$ the two kinds of models are indistinguishable
    because they both depend on a single parameter, $\nu$ and $\nueff$
    respectively.
\end{itemize}

Let us next proceed with the numerical fit analysis. In order to place
tighter constraints on the corresponding parameter space of our model, the
probes described above must be combined through a joint likelihood
analysis\footnote{Likelihoods are normalized to their maximum values. In
the present analysis we always report $1\sigma$ uncertainties on the
fitted parameters. Note also that the total number of data points used
here is
$N_{tot}=559$, while the associated degrees of freedom is: {\em d.o.f}$\
\, = N_{tot}-n_{\rm fit}-1$, where $n_{\rm fit}$ is the model-dependent
number of fitted parameters.}, given by the product of the individual
likelihoods according to:
\begin{equation}\label{eq:overalllikelihood} {\cal L}_{\rm tot}({\bf p})=
{\cal L}_{\rm SNIa} \times {\cal L}_{\rm BAO} \times {\cal L}_{\rm
CMB}\;, \end{equation}
Since likelihoods are defined as ${\cal L}_j\propto
\exp{\left(-\chi_j^2/2\right)}$, it translates into an addition for the
joint $\chi^2$ function:
\begin{equation}\label{eq:overalllikelihoo}
\chi^{2}_{\rm tot}({\bf p})=\chi^{2}_{\rm SNIa}+ \chi^{2}_{\rm
BAO}+\chi^{2}_{\rm CMB}\;.
\end{equation}
In our $\chi^2$ minimization procedure, for the vacuum models (running and
concordance $\Lambda$CDM) we use the following range and steps for the
fitting parameters: $\tilde{\Omega}_{m}^{0} \in [0.01,1]$ in steps of 0.001 and
$\nueff \in [-0.02,0.02]$  in steps of $10^{-4}$.
%
\begin{figure}[t]
\hspace{0cm}\phantom{XXXXXX}{$\nueff$}\\
\includegraphics[scale=0.9]{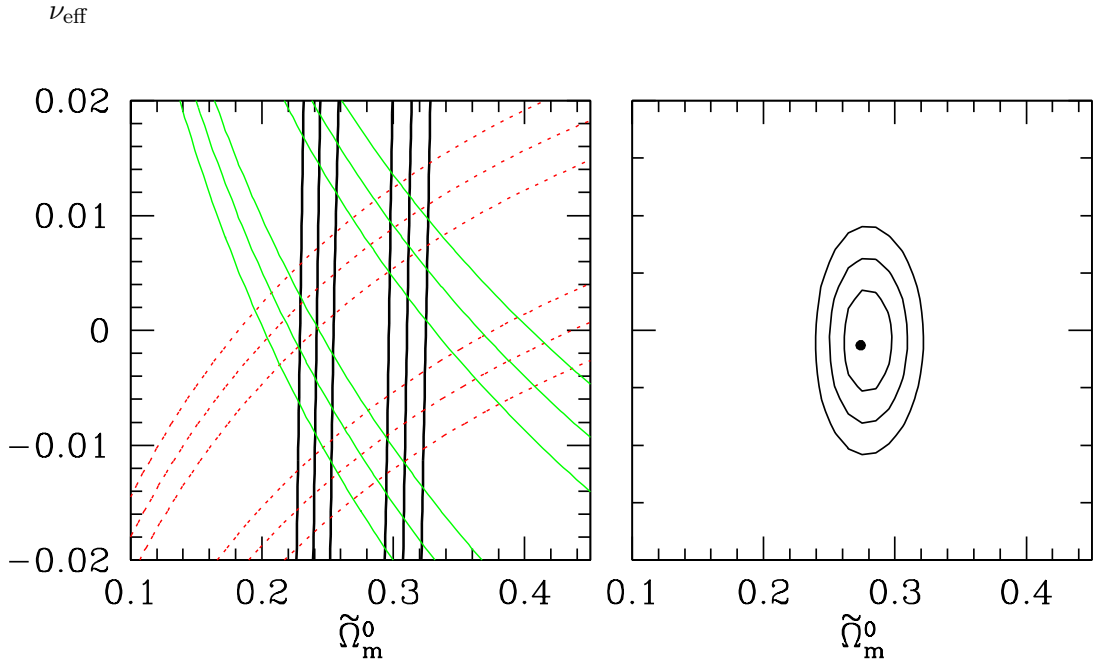}
\caption{\footnotesize{Likelihood contours (for $-2{\rm ln}{\cal
L}/{\cal L}_{\rm max}$ equal to 2.30, 6.16 and 11.81, corresponding
to 1$\sigma$, 2$\sigma$ and $3\sigma$ confidence levels) in the
$(\tilde{\Omega}_{m}^{0},\nueff)$ plane
for the generalized running
vacuum model (\ref{HubbleFlow3}) ($C_0\neq 0$ or $\tilde{\Omega}^0_{\Lambda}\neq 0$).
For the CMB analysis we
include also the radiation component as indicated in (\ref{mixturev})-(\ref{HubbleFlow3}). The left
panel shows the contours based on the SNIa data (thick solid black
lines), BAOs (dotted-red lines) and CMB shift parameter
(dashed-green lines). In the right panel we show the corresponding
contours based on the joint statistical analysis of the SNIa+BAO+CMB
data.}\label{fig:contours}}
\end{figure}
The numerical results that we obtain are the following. In the case of the
generalized running vacuum model the overall likelihood function peaks at
$\tilde{\Omega}_{m}^{0} = 0.274\pm 0.011$, $\nueff=-0.00133\pm 0.0028$ (or
$\nu=4\nueff \simeq -0.00532$, $\alpha=3\nueff\simeq-0.004$) with
$\chi_{\rm tot}^{2}(\tilde{\Omega}_{m}^{0},\nueff) \simeq 542.93$ for
$556$ degrees of freedom\footnote{Note that in \cite{GSBP11} the original
running vacuum model was used -- see Eqs.\,(\ref{RGlaw2}) and
(\ref{vacuuma}) --, in which $\alpha$ is strictly equal to zero -- and the
{\em Constitution} set of 397 SNIa data\,\cite{Hic09}. We would like to
mention here that those results for $\nu$ are in agreement with the
current results for $\nueff$ within $1\sigma$ uncertainties.}.
%
%
In Fig.\,\ref{fig:contours} we present the 1$\sigma$, 2$\sigma$ and
$3\sigma$ confidence levels in the $(\tilde{\Omega}_{m}^{0},\nueff)$
plane. In particular, the left panel in that figure shows the individual
likelihood contours, with the SNIa-based results indicated by thick solid
lines, the BAO results by dotted-red lines and those based on the CMB
shift parameter by dashed-green lines. Using the SNIa data alone it is
evident that although the $\tilde{\Omega}_{m}^{0}$ parameter is tightly
constrained ($\simeq 0.27$), the $\nueff$ parameter remains completely
unconstrained. As can be seen in the right plots of Figure 1, the above
degeneracy is broken when using the joint likelihood analysis, involving
all the cosmological data. Finally, in the case of the concordance
$\Lambda$CDM cosmology ($\nueff=\alpha\equiv 0$) we find $\OMo=0.274\pm
0.01$ with $\chi_{\rm tot}^{2}(\OMo)/d.o.f \simeq 543.18/558$.

Overall we see that the departure of the GRVE model with respect to the
$\CC$CDM is extremely small and cannot be detected at present.
%

%
\mysection{Discussion and conclusions} \label{sect:conclusions}

In this paper we have generalized the running vacuum energy  models and we
have solved the corresponding background cosmology. The generalized
running vacuum model (\ref{FR1})-(\ref{GRVE}) with  $C_0\neq 0$ is able to
pass the SNIa+BAO+CMB data constraints with a statistical significance
comparable to that of the concordance $\Lambda$CDM model which is a
limiting case of the model ($\xi_r=\xi_m=1$, or equivalently
$\alpha=\nu=0$). Although the best-fit models are currently
indistinguishable from $\Lambda$CDM we expect that future very accurate
data on both low and high redshifts could help to distinguish these models
from the standard cosmology.

Some conceptual issues pertaining to these models were also addressed
which are related to the peculiar conservation laws derived in Section 3.
We stress that these issues hold as well for other models with analogous
effective Friedmann equations, models inspired either by the holographic
or the entropic-force principle. We have further emphasized that the
presence of a non vanishing additive constant $C_0$ is crucial since
otherwise the cosmology does not allow for a transition between decelerated
and accelerated expansion. That was actually noticed in previous entropic-force
studies \cite{Koivisto10}\,\footnote{For other problems related with the
entropic-force cosmology, see the recent\, \cite{RovetoMunoz12}.}. In
contrast, the class of the running vacuum models, both the generalized one
(GRVE) presented here and the original one (which existed in the literature
since long ago -- see \,\cite{MiniReview11} and references therein)  do not
suffer from this problem because $C_0$ is naturally expected to be non-vanishing
as a result of integrating the corresponding RG equation. Therefore,
despite the formal analogies between these two sorts of models, the
running vacuum models are naturally well positioned for a correct
phenomenological description of our cosmos.

From the point of view of the running vacuum models, the current Universe
appears as FLRW-like with a genuine cosmological constant while dust and
radiation evolve in a nonstandard way, in the sense that they follow
scaling laws that deviate slightly from their behavior in $\CC$CDM. In
contrast to the old running vacuum energy model, the generalized one
introduced in this paper allows for an independent departure from the
standard behavior of both components. We have used this freedom and we
have explored models satisfying $\xi_r=1$ thereby ensuring that
relativistic matter obeys the standard behavior. In this way potential
difficulties related to the radiation dominated era are essentially
avoided. While the other parameter $\xi_m\simeq 1-\nueff$ remains free, it
can be efficiently constrained using CMB data. It is constrained by
observations at a similar level as the single parameter $\nu$ of the
original running model, i.e. they are both presently allowed up to ${\cal
O}(10^{-3})$ at most (in absolute value).

This order of magnitude size is consistent with the
theoretical expectations on these coefficients, interpreted as one-loop
$\beta$-functions of the running cosmological constant. The mild variation
induced on the CC term by these coefficients is responsible for the
dynamical character of the vacuum energy, which is of course the reason
why these models have a chance to improve the situation with the $\CC$CDM
without giving up its phenomenological success. Such time variation is
foreseen on general QFT grounds and it provides a possible formulation of
an effective dynamical dark energy, which in some cases can help curing
the cosmic coincidence problem\,\cite{LXCDMperturbations} and other related
problems.

To summarize, the running vacuum models offer a challenging
phenomenologically consistent description of a universe with presently
accelerated expansion. The dynamical $\CC$ could be understood in the
context of QFT in curved space-time. Such potential connection with
fundamental physics could help to conceive the origin of a dynamical $\CC$
term in QFT and eventually provide an explanation for the tough cosmological
constant problem.

\vspace{0.5cm}

{\bf Acknowledgments} \vspace{0.2cm}

JS has been supported in part by MEC and FEDER under PA2010-20807, by the
Spanish program CPAN CSD2007-00042 and by 2009SGR502 Generalitat de
Catalunya. SB thanks the Dept.\ ECM of the Univ.\ de Barcelona for the
hospitality, and the financial support from the Spanish Ministerio de
Education, within the project SAB2010-0118.

\newcommand{\JHEP}[3]{ {JHEP} {#1} (#2)  {#3}}
\newcommand{\NPB}[3]{{ Nucl. Phys. } {\bf B#1} (#2)  {#3}}
\newcommand{\NPPS}[3]{{ Nucl. Phys. Proc. Supp. } {\bf #1} (#2)  {#3}}
\newcommand{\PRD}[3]{{ Phys. Rev. } {\bf D#1} (#2)   {#3}}
\newcommand{\PLB}[3]{{ Phys. Lett. } {\bf B#1} (#2)  {#3}}
\newcommand{\EPJ}[3]{{ Eur. Phys. J } {\bf C#1} (#2)  {#3}}
\newcommand{\PR}[3]{{ Phys. Rep. } {\bf #1} (#2)  {#3}}
\newcommand{\RMP}[3]{{ Rev. Mod. Phys. } {\bf #1} (#2)  {#3}}
\newcommand{\IJMP}[3]{{ Int. J. of Mod. Phys. } {\bf #1} (#2)  {#3}}
\newcommand{\PRL}[3]{{ Phys. Rev. Lett. } {\bf #1} (#2) {#3}}
\newcommand{\ZFP}[3]{{ Zeitsch. f. Physik } {\bf C#1} (#2)  {#3}}
\newcommand{\MPLA}[3]{{ Mod. Phys. Lett. } {\bf A#1} (#2) {#3}}
\newcommand{\CQG}[3]{{ Class. Quant. Grav. } {\bf #1} (#2) {#3}}
\newcommand{\JCAP}[3]{{ JCAP} {\bf#1} (#2)  {#3}}
\newcommand{\APJ}[3]{{ Astrophys. J. } {\bf #1} (#2)  {#3}}
\newcommand{\AMJ}[3]{{ Astronom. J. } {\bf #1} (#2)  {#3}}
\newcommand{\APP}[3]{{ Astropart. Phys. } {\bf #1} (#2)  {#3}}
\newcommand{\AAP}[3]{{ Astron. Astrophys. } {\bf #1} (#2)  {#3}}
\newcommand{\MNRAS}[3]{{ Mon. Not. Roy. Astron. Soc.} {\bf #1} (#2)  {#3}}
\newcommand{\JPA}[3]{{ J. Phys. A: Math. Theor.} {\bf #1} (#2)  {#3}}
\newcommand{\ProgS}[3]{{ Prog. Theor. Phys. Supp.} {\bf #1} (#2)  {#3}}
\newcommand{\APJS}[3]{{ Astrophys. J. Supl.} {\bf #1} (#2)  {#3}}

\newcommand{\Prog}[3]{{ Prog. Theor. Phys.} {\bf #1}  (#2) {#3}}
\newcommand{\IJMPA}[3]{{ Int. J. of Mod. Phys. A} {\bf #1}  {(#2)} {#3}}
\newcommand{\IJMPD}[3]{{ Int. J. of Mod. Phys. D} {\bf #1}  {(#2)} {#3}}
\newcommand{\GRG}[3]{{ Gen. Rel. Grav.} {\bf #1}  {(#2)} {#3}}



\end{document}